\documentclass[10pt, journal, twocolumn]{IEEEtran}

\usepackage[utf8]{inputenc}		
\usepackage[scr=boondoxo,frak=euler,bb=ams]{mathalfa}
\usepackage{amsmath,amsthm,amsfonts,amssymb} 	
\usepackage{graphicx,graphics}
\usepackage{float}
\usepackage{tikz}
\usetikzlibrary{shapes, snakes, patterns}
\usepackage{xcolor}		

\usepackage[export]{adjustbox}
\usepackage[justification=centering]{caption}
\usepackage{subcaption}
\DeclareCaptionLabelSeparator{periodspace}{.\quad}
\usepackage{epstopdf}
\usepackage{enumerate,enumitem} 
\usepackage{cite}
\usepackage{suffix}
\usepackage{mathtools}
\usepackage{tabularx}
\usepackage{booktabs}		
\usepackage{boldline,multirow,diagbox,hhline} 
\usepackage{slashbox}
\usepackage{algorithm,algpseudocode}
\usepackage{relsize}
\usepackage{cuted}
\usepackage{titlesec}
\usepackage{chngcntr}
\usepackage{textcomp}
\def\BibTeX{{\rm B\kern-.05em{\sc i\kern-.025em b}\kern-.08em
		T\kern-.1667em\lower.7ex\hbox{E}\kern-.125emX}}

\theoremstyle{definition}

\newtheorem{definition}[]{Definition}

\newtheorem{remark}[]{Remark}

\newcommand{\PreserveBackslash}[1]{\let\temp=\\#1\let\\=\temp}
\newcolumntype{C}[1]{>{\PreserveBackslash\centering}p{#1}} 
\newcolumntype{R}[1]{>{\PreserveBackslash\raggedleft}p{#1}} 
\newcolumntype{L}[1]{>{\PreserveBackslash\raggedright}p{#1}} 

\newcolumntype{Y}{>{\centering\arraybackslash}b{0.7cm}}
\newcolumntype{M}{>{\centering\arraybackslash}b{2.1cm}}
\newcolumntype{Z}{>{\centering\arraybackslash}b{1.75cm}}
\newcolumntype{G}{>{\centering\arraybackslash}m{1.5cm}}
\newcolumntype{Q}{>{\centering\arraybackslash}b{3.75cm}}
\newcolumntype{A}{>{\centering\arraybackslash}m{1.5cm}}
\newcolumntype{B}{>{\centering\arraybackslash}m{0.6667cm}}
\newcolumntype{T}{>{\centering\arraybackslash}b{6.5ex}}
\newcolumntype{S}{>{\centering\arraybackslash}b{13.25ex}}
\newcolumntype{?}{!{\vrule width 1pt}}
\newcolumntype{+}{!{\vrule width 2pt}}

\DeclareFontFamily{U}{mathx}{}
\DeclareFontShape{U}{mathx}{m}{n}{<-> mathx10}{}
\DeclareSymbolFont{mathx}{U}{mathx}{m}{n}
\DeclareMathAccent{\widecheck}{0}{mathx}{"71}

\definecolor{bluee}{RGB}{0, 82, 200}
\definecolor{redd}{RGB}{245, 30, 30}
\definecolor{greenn}{RGB}{0, 150, 62}

\newcommand{\mbf}[1]{\boldsymbol{\mathrm{#1}}}

\newcommand{\abs}[1]{\left\lvert#1\right\rvert}
\newcommand{\imag}{\mathrm{j}}

\newcommand{\twiddle}[1]{\text{\small{W}}_{\nsp \text{\scriptsize{\textit{{N}}}}}^{ #1 {}^{\phantom1}_{}}\!\!\nspp}

\newcommand{\commreplace}[2]{\textcolor{blue}{#1}} 
\newcommand{\hyp}{$\pspp$-$\pspp$}

\newcommand{\commhide}[1]{}

\newcommand{\psp}{\hspace{0.1em}}
\newcommand{\pspp}{\hspace{0.05em}}
\newcommand{\nsp}{\hspace{-0.1em}}
\newcommand{\nspp}{\hspace{-0.05em}}

\makeatletter
\newcommand{\vast}{\bBigg@{4}} 
\newcommand{\Vast}{\bBigg@{5}} 
\newcommand{\csize}[1]{\bBigg@{#1}} 
\makeatother

\makeatletter
\newcommand{\ostar}{\mathbin{\mathpalette\make@circled\star}}
\newcommand{\make@circled}[2]{\ooalign{$\m@th#1\smallbigcirc{#1}$\cr\hidewidth$\m@th#1#2$\hidewidth\cr}}
\newcommand{\smallbigcirc}[1]{ \vcenter{\hbox{\scalebox{0.77778}{$\m@th#1\bigcirc$}}}}
\makeatother

\newcommand{\conj}[1]{{#1}^{*}_{}}

\newcommand{\modulo}[2]{\langle {#1} \rangle_{#2}^{}}
\newcommand{\bigmodulo}[2]{\big\langle {#1} \big\rangle_{#2}^{}}

\title{Using DCFT for Multi-Target Detection in Distributed Radar Systems with Several Transmitters}

\author{
	\IEEEauthorblockN{%
		Gokularam Muthukrishnan, %
		$\,\ $%
		S. Sruti, %
		$\,\ $%
		K. Giridhar\\}
	\IEEEauthorblockA{\textit{TelWiSe Group, Department of Electrical Engineering,} \\
		\textit{Indian Institute of Technology Madras,}\\
		Chennai - 600036, 
		India.\\
		e-mail: \{gokularam, srutisiva, k.giridhar\}@telwise-research.com}
}

\begin{document}
	
	\maketitle	
	
	\begin{abstract} 
		%
		%
		%
		In distributed radar systems, when several transmitters radiate simultaneously, 
		the reflected signals 
		need to be distinguished at the receivers to detect various targets. 
		If the transmit signals are in different frequency bands, they require a large overall bandwidth. 
		%
		Instead, a set of pseudo-orthogonal waveforms derived from the Zadoff-Chu (ZC) sequences 
		could be 
		accommodated in the same band, 
		enabling the efficient use of available bandwidth for better range resolution. 
		%
		In such a design, special care must be given to the `near-far' problem, where a reflection could possibly become difficult to detect due to the presence of stronger reflections. 
		%
		%
		%
		%
		%
		%
		In this work, a scheme to detect multiple targets in such distributed radar systems 
		is proposed. 
		It performs successive cancellations (SC) starting from the strong, detectable reflections in the domain of the Discrete Chirp-Fourier Transform (DCFT) after compensating for Doppler shifts, 
		enabling the subsequent detections of weaker targets which are not trivially detectable. 
		Numerical 
		simulations corroborate the efficacy and usefulness 
		of the proposed method in detecting weak target reflections. 
	\end{abstract}
	
	
	\section{Introduction}\label{sec:intro}

	Distributed Radar Systems (DRS)%
	\cite{chernyak2018fundamentals,li2008mimo,nguyen2019signal,deng2020radar}, unlike conventional phased array-based monostatic radars,
	use multiple antennas at separate geographical locations for the transmission and reception of radio waves 
	with the aim of detecting, localizing and tracking air-borne targets in the area of interest. 
	When the transmitters and receivers are widely distributed such that the separations between them form a significant fraction of the target range, it is possible to collect the obliquely scattered energy from the targets	at different aspect angles, and this spatial diversity can be exploited to improve the performance \cite{haimovich2007mimo}. 
Observing radar cross-section (RCS) signatures of targets at different angles 
remarkably improves detection performance \cite{li2015moving} and enhances 
target localization \cite{godrich2010target} and velocity estimation \cite{noroozi2020efficient} accuracy.
DRS also  
enables the detection of slow-moving targets 
and stealth targets 
and offers covertness 
and immunity to electronic counter-measures. It even facilitates imaging and classification of targets \cite{sruti2023rcs}.	

Two kinds of DRS exist: the passive and the active. 
Passive DRS
operates with `illuminators of opportunity' \cite{kuschel2019tutorial}; several types of transmitters of opportunity shall be used for the passive operation, e.g.,  DVB-T, DVB-S, DAB, FM, communication links or even existing monostatic radars.
However, such systems 
exploit waveforms that are not necessarily good for target detection, 
resulting in poor performance despite the complex signal processing at the receivers. 
%
%
In this paper, we focus on an active, continuous wave (CW) DRS, which employs several dedicated 
transmitters that continuously radiate 
repetitions of a specific wave pattern 
that are known to the receivers. 
%
These waveforms are reflected back from the prospective targets toward the receive antennas, which can then be processed and fused at various levels.
%
%
%
%

A received waveform after impinging a target would be delayed, attenuated and also shifted in frequency if the target is in motion. 
%
%
%
%
The time delay (TD) and Doppler shift (DS) are the key signatures 
that a target leaves in a waveform. 
TD corresponds to the total distance travelled by the waveform (bistatic range, BR), whose rate of change (bistatic range rate, BRR) translates to DS in the waveform.
Once a target is detected, estimated TDs and DSs of the reflected waveforms at various receivers shall be 
fused to estimate its kinematic state, i.e., position and velocity 
\cite{noroozi2020efficient}. 
At the \textit{matched filter} receivers that operate non-coherently, targets are first detected, and the corresponding TD and DS are then estimated by computing the cross ambiguity function between received and reference transmit signals.

For good performance, waveforms must have tolerance to Doppler and low autocorrelation side lobes. 
%
%
Also, it is desirable to have a low Peak to Average Power Ratio (PAPR) for efficient transmission. 
The use of periodic modulated CW signals can yield perfect autocorrelation \cite{saunders1990cw}, and several spreading codes have been studied in the literature for this purpose \cite{levanon2004radar,benedetto2008role}. 
%
%
Constant Amplitude Zero Auto-Correlation (CAZAC) sequences exhibit ideal time localization and transmission efficiency.
Certain classes of CAZAC sequences have been used in radars for years. 
%
%
Among the CAZAC sequences, Zadoff–Chu (ZC) \cite{chu1972polyphase} polyphase codes are a good choice here because of their flexible length and robustness against Doppler effects \cite{levanon2004radar}. 
%
%
ZC sequences are \textit{perfect sequences} with an ideal periodic autocorrelation function (PACF) 
and have unit PAPR. 
Periodic cross-correlation function (PCCF) between two ZC codes from different seeds is also low 
and documented in \cite{kang2011generalized}.
%
%
%
%
%
%

The use of multiple transmitters offers robust, fault-resilient functioning of DRS 
even if some transmitters are compromised. 
Also, having more transmitters improves the coverage and provides better localization accuracy \cite{noroozi2015target}. 
However, it has its own set of challenges;
when the receiver collects echoes from multiple transmitters, they have to be uniquely identified and separately processed to leverage the benefits. Thus, mutual interference between the different transmit signals should be minimized.
%
Transmitters can operate orthogonally in frequency with no bandwidth overlap or in time without simultaneous radiations.
The accuracy of 
TD measurements is limited by the bandwidth, and hence, the  
accuracy of the target position estimates depends strongly on the bandwidth 
\cite{godrich2010target}.
The paucity of available spectrum 
%
renders	the sharing of available bandwidth among the transmitters difficult without compromising localization accuracy. 
Longer integration times are required for good Doppler resolution. Also, longer signal duration results in better accuracy of the target speed estimation \cite{he2010target}; consequently, 
switching 
between the available transmitters before the targets have significantly moved would result in poor performance.
%
%
%
%
\commreplace{\\{}\indent$\!\!\!$}{}
By leaving these options, waveforms are necessitated to be mutually orthogonal under correlation  (i.e., orthogonal in code) 
while transmitted simultaneously over the same spectrum.
A set of waveforms with an impulsive autocorrelation function and zero mutual cross-correlation functions is desirable. However, it is impossible to design such an ideal set of sequences  \cite{sarwate1979bounds,zepernick2013pseudo}.
However, to the rescue, polyphase codes,  such as the ZC codes, have good cross-correlation properties alongside ideal PACF%
\cite{deng2004polyphase}. Prime length ZC codes attain the tightest possible suppression by a factor of $\sqrt{{N}}$ for perfect sequences, where ${N}$ is the code length; 
for other cases, there is only a graceful degradation \cite{kang2011generalized}. For these reasons, we consider ZC sequences for the  active CW  DRS with several transmitters.

When there are several targets  in the observed space, each one of them needs to be detected, localized and tracked separately. 
However, when the targets are at different heights, the power of the echoed transmit waveforms will have large variations, and some targets might  go undetected even with good PCCF. 
Reflections from targets which are high or which have low RCS will be weak; 
correlation side lobes arising from the strong reflections from certain targets might mask such weak target echoes, rendering them undetectable and, consequently, limiting the range of observable space. 
%
Thus, appropriate processing techniques to detect need to be designed such that weak reflections from targets can be identified when there are strong reflections from other targets.
In this work, we address this problem.
%
%
%
\subsection{Motivation}
%
%
%
%
%
Several signal processing techniques in DRS draw inspiration from those in MIMO wireless communication systems 
\cite{li2008mimo} and one such idea, namely the successive cancellation (SC) \cite{wubben2001efficient,vanka2012superposition}, is used in this work for multi-target detection in active CW DRS with several transmitters. It can be observed that ZC sequences resemble samples of 
linear frequency modulated (LFM) waveforms (or chirps) used in conventional radars, and the seed of a sequence determines the sweep rate of the frequency. Discrete Chirp-Fourier Transform (DCFT) 
\cite{xia2000discrete} is a generalization of Discrete Fourier Transform (DFT) \cite{oppenheim1999discrete} that 
can reveal the presence of chirps when a variable in the transform is matched to the sweep rate. Consequently, a ZC sequence present in the received composite signal can be transformed to an impulse in the transform domain by matching the chirp rate variable to the seed of the sequence, where it can be filtered by nulling the impulse. Since this transform is invertible, the ZC sequence would have been excised in the time domain after 
this step, 
leaving room  to detect any weaker ZC sequence present in the residue. By appropriately handling the Doppler in the reflected waveforms, this process can be repeated until no more detectable target reflections are present in the residue. 
This idea for multi-target detection is formally presented in Section \ref{sec:algorithm}.





\subsection{Basic Notations}
In this article, $\imag$ indicates the imaginary unit, i.e., $\imag=\sqrt{-1}\pspp$. 
The set of all complex numbers, the set of all integers, and the set of all positive integers are denoted as $\mathbb{C}\pspp$, $\mathbb{Z}$ and $\mathbb{N}\pspp$, respectively. 
The complex conjugate of $v\in\mathbb{C}$ is written as $\conj{v}\nsp\pspp$. 
$\operatorname{gcd}({a},{b})$ 
is 
the greatest common divisor of 
${a}\in \mathbb{Z}$ and ${b}\in \mathbb{Z}$ 
and is 
non-negative by convention; 
if $\operatorname{gcd}({a},{b})=1\pspp$, $a$ and $b$ are said to be relatively prime, and $\operatorname{gcd}(a,0)=\abs{a}$. 
%
The modulo operator
$\modulo{{a}}{{b}}$ 
provides the (non-negative) remainder of the Euclidean division of 
${a}\in \mathbb{Z}$ by ${b}\in \mathbb{Z}\backslash\{0\}\pspp$; hence, $\modulo{{a}}{{b}}\in\{0,\,1,\,\ldots,\,\abs{{b}\hspace{0.035em}}-1\}\pspp$.
\commreplace{\\{}\indent$\!\!\!$}{}
%
The Kronecker delta	function 
is denoted as $\delta[\pspp\cdot\pspp]\pspp$. Consider two complex-valued ${N}$-length sequences, $x[n]$ and $y[n] \pspp$, $n=0,\,1,\,\ldots,\,{N}-1$. The periodic correlation function between them at the shift $l \in \{0,\,1,\,\ldots,\,{N}-1\}$ is evaluated as 
$(x\ostar y)[\pspp l \pspp]=\sum_{n=0}^{{N}-1} x[n] \psp \conj{y}[\modulo{n-l}{{N}}]
\pspp$; if both the sequences are the same, it is called periodic autocorrelation function (PACF), else periodic cross-correlation function (PCCF).

\section{Background and System Model}\label{sec:background}

Let us take a short digression to review the necessary background and introduce the system model.

\subsection{Zadoff-Chu Sequences}


Zadoff-Chu (ZC) sequences \cite{chu1972polyphase} are complex-valued codes that are widely used in various signal processing applications%
\cite{gul2014timing,piccinni2017novel,lee2013zadoff}. 
They are constant amplitude, polyphase sequences whose elements 
are roots of unity. 

\begin{definition} \label{defn:zc}
	The ZC 
	sequence of length ${N}$ 
	is given by
	%
	%
	\begin{equation} \label{eq:ZC}
		z_{u,{N}}^{}[n] = \twiddle{u\pspp{n ( n+\modulo{{N}}{2} ) } / {2}}
		\pspp,
		\,\ n=0,\,1,\,\ldots,\,{N}-1
		\psp,
	\end{equation}
	where $\twiddle{} \,=e^{-\imag\pspp{2\pi}/{{N}}}_{}
	$ 
	denotes the well-known \textit{twiddle factor}, 
	%
	%
	and $u\in\mathbb{Z}\pspp$, known as the \textit{seed} of the sequence, is relatively prime to ${N}\nspp\pspp$.
\end{definition}

\noindent Thus, the seed-$u$ ZC sequence is a unique 
%
%
%
%
code comprising a particular arrangement of ${N}$-th roots of unity arising from the choice of primitive 
root $\twiddle{u}\nspp\pspp$. 
%
%
%
%
%

ZC sequences have several nice 
properties.
It can be observed that any ZC sequence is periodic with period ${N}$, i.e., $z_{u,{N}}^{}[n]=z_{u,{N}}^{}[\modulo{n}{{N}}] \,\ \forall\, n\in\mathbb{Z}$; thus, Definition \ref{defn:zc} holds for all $n\in\mathbb{Z}\pspp$. 
Also, 
a time delay in a ZC sequence manifests as a Doppler shift to the  original waveform, i.e., 
$z_{u,N}^{}[n-l]=
z_{u,N}^{}[n] 
\times
z_{u,N}^{}[-l] \psp 
\pspp\twiddle{-uln}\nspp\pspp
$; thus, ZC sequences exhibit time-frequency coupling. 
%
%
%
%
%
%
%
%

The DFT or IDFT of a ZC Sequence is again a ZC Sequence \cite{li2007constructive}. 
Also, ZC sequences have desirable correlation characteristics, which are listed below, making them well-suited for radar and communication applications.
\begin{enumerate}[label=(\roman*)]
	
	\item 
	ZC sequences are perfect sequences, and they are orthogonal to cyclically shifted copies of themselves, irrespective of the seed \cite{chu1972polyphase}. Thus, they have an ideal 
	PACF
	of $N\pspp\delta[\modulo{\pspp l\pspp }{N}]\pspp$.
	
	\item 
	The maximum absolute value of the 
	PCCF 
	between the length-${N}$ ZC sequences arising from two different seeds, ${a}$ 
	and ${b}\pspp$, 
	is given by  $\sqrt{{\tau} {N}}$, where ${\tau}=\operatorname{gcd}({N},{a}-{b})$ 
	\cite{kang2011generalized}. 
\end{enumerate}
Variations such as cyclic shifts, the addition of a constant phase, or conjugating the entire code 
will not affect the PACF characteristics of the ZC sequences.
\commreplace{\\{}\indent$\!\!\!$}{}
%
ZC codes can be 
interpreted as discrete counterparts 
of linear frequency modulated (LFM) signals (or chirps); 
the sweep rate of the frequency is determined by the seed of the ZC sequence.  
Subsequently,  their energies are concentrated as (aliased) 
\textit{lines} in the time-frequency plane%
, as observed in the normalized spectrogram of a ZC sequence in Figure \ref{fig:zc_spectrogram}. 
This interpretation makes the use of 
signal processing
tools available for 
chirps	to ZC sequences possible: we witness one such instance in the sequel through DCFT.

\commreplace{
	\begin{figure}[ht]
		\begin{center}
			\centerline{\includegraphics[width=0.92\columnwidth]{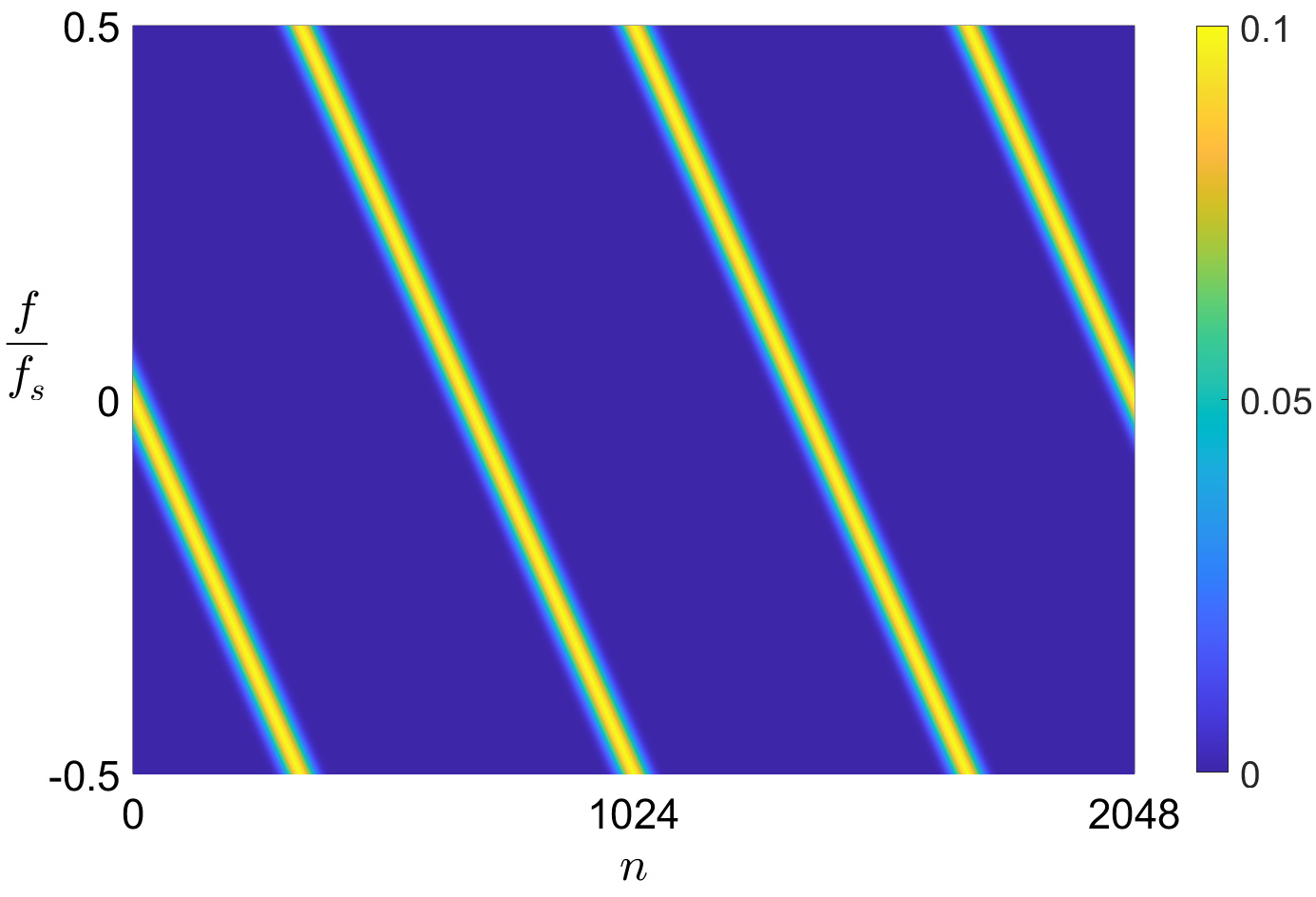}}
			\caption{Normalized spectrogram of Zadoff-Chu sequence of length $N=2048$ with seed $u=3$.}
			\label{fig:zc_spectrogram}
		\end{center}
		\vskip -0.3ex
	\end{figure}
}{
}


\subsection{Discrete Chirp-Fourier Transform}

Following the observation that the ZC sequences are the sampled chirp signals,  we now introduce a tool for filtering them in the receivers. Classical Fourier spectrum analysis tools like DFT  become useless for chirps  as they 
cannot be used to analyze wideband and non-stationary signals. 
The Discrete Chirp-Fourier Transform (DCFT) \cite{xia2000discrete}
generalizes the DFT to facilitate the analysis of chirp-type signals,  
and it comes in handy in radar applications. 
DCFT has been used for 
detection \cite{li2017multicomponent} and  parameter estimation \cite{yang2015parameter} of multi-component chirp signals, 
Detection for high-speed manoeuvring targets \cite{huang2019ground}
and
identification of jamming \cite{wu2020recognition}%
.
\commreplace{\\{}\indent$\!\!\!$}{}
DCFT is parametrized by the chirp rate variable ${\beta}$ and the 
frequency index 
${k}\pspp$. 
DCFT is related to Fractional Fourier Transform (FrFT) \cite{almeida1994fractional}, and the variable $\beta$ in the DCFT corresponds to the rotation angle in FrFT. DCFT and its inverse are defined as follows\footnote{We consider the modified DCFT, MDCFT I in \cite{fan2001two}, with a slight change.}.

\begin{definition}[DCFT and IDCFT]\label{defn:dcft}
	The ${N}$-point DCFT%
	\commreplace{}{\footnote{We consider the modified DCFT, MDCFT I in \cite{fan2001two} with a slight change.} }
	of an ${N}$-length sequence $x[n]\pspp$, $n=0,\,1,\,\ldots,{N}-1$ is given by
	\begin{equation}\label{eq:dcft}
		\begin{split}
			&
			X_{\texttt{{c}}}^{}[{k},{\beta}]
			=\frac{1}{\sqrt{{N}}} \sum_{n=0}^{N-1} x[n]\, \twiddle{kn-{\beta}n^2/2}
			,
			\\& \hspace{3.55em}
			{k}=0,\,1,\,\ldots,\,{N}-1\pspp,
			\,\
			{\beta}=0,\,1,\,\ldots,\,\big\lceil\tfrac{{N}}{2}\big\rceil-2
			\pspp
			\commreplace{.}{,}
		\end{split}\!\!
	\end{equation}
	Thus, $\{X_{\texttt{{c}}}^{}[{k},{\beta}]\}_{{k}=0}^{{N}-1}$ is the DFT of $x[n]\,\twiddle{-{\beta}n^2/2}$ for an arbitrarily fixed ${\beta}\in\big\{0,\,1,\,\ldots,\,\big\lceil\tfrac{{N}}{2}\big\rceil-2\big\}\pspp$, and the inverse DCFT (IDCFT) can be defined as
	\begin{equation}\label{eq:idcft}
		x[n]
		= 
		\frac{1}{\sqrt{{N}}} \sum_{{k}=0}^{N-1} X_{\texttt{{c}}}^{}[{k},{\beta}]\, \twiddle{-{k}n+{\beta}n^2/2}
		,
		\,\ n=0,\,1,\,\ldots,\,{N}-1
		\pspp.
	\end{equation}
\end{definition}
\noindent Thus, the DCFT 
reduces to the DFT when ${\beta}=0\pspp$. 

\begin{remark}
	For the remainder of 
	this paper, we will consider only even length ZC sequences ${N}\nspp\pspp$, ${N}\in2\mathbb{N}
	\pspp$, particularly ${N}$, which is a power of 2.  
	This  is mainly to facilitate the implementation of DCFT and PCCF via the Fast Fourier Transform (FFT) algorithm  with lesser computations. 
	In such cases, any odd integer ${u}$ will be 
	relatively prime to
	${N}$. The seed $u$ is further restricted to be in the set 
	$u\in\{ 1,\,2,\,\ldots,\, {N}/2-2\}$ in order to unambiguously match 
	${\beta}$ and ${u}\pspp$.
\end{remark}

\subsection{System Model}

Let us consider an active CW DRS comprising of ${M}$ widely separated transmit antennas, each of which periodically emits an ${N}$-length ZC waveform of unique seed, and  the reflections of these waveforms from the ${K}$ targets present in the observation area are collected by ${L}$ receive antennas%
, themselves widely placed.
%
%
The samples of the transmit signal are given by 
\begin{equation}
	{s}_i^{}[n]=z_{u_i^{},N}^{}[n], 
	\quad i=1,\,2,\,\ldots,\,M
	\pspp,
\end{equation}
where $u_i^{}\in\{ 1,\,\ldots,\, {N}/2-2\}\pspp$, $u_i^{}\neq u_m^{}$ for $i\neq m$ and $\gcd(u_i^{},N)=1\,\ \forall\, i\in1,\,2,\,\ldots,\,M\nspp\pspp$. 
Optimal selection of ZC codes for transmitters meeting these constraints can be done by picking seeds based on the results presented in \cite{kang2011generalized}. 
%
%
However, our numerical studies indicated that these choices did not impact the performance much,  
and therefore, we simply picked the first ${M}$ odd integers as seeds, i.e., $ u_i^{}=2i-1, \quad i=1,\,2,\,\ldots,\,M\pspp$.
%
%
%
%
%
%

In this work, we assume that all the ${L}$ targets are point targets, occupying a single range-Doppler bin at the receivers' matched filters at most.
Also, the targets are assumed to move approximately at constant velocities over the coherence interval.
It is assumed that there is no line of sight between the transmitters and the receivers, and the direct signal from any transmitter is not observed by the receivers through the use of appropriate antenna beam patterns. 
We also assume that there is no clutter since the targets are air-borne.
%
%

Each receiver collects the signals reflected by the targets. 
The samples of the received baseband signals are modelled as
\begin{equation}\label{eq:rx}
	\begin{split}
		{v}_p^{}[n] &= {v}_{p}^{(0)}[n] + {\gamma}_p^{}[n], \quad p=1,\,2,\,\ldots,\,{L}, \,\ \text{ where }
		\\
		{v}_{p}^{(0)}[n] &= \sum_{i=1}^{M} \sum_{q=1}^{K} \alpha_{ip}^{(q)} \pspp 
		s_i^{}\big[n-l_{ip}^{(q)}\big] \pspp
		\exp\nsp\Big(\imag 2\pi {\xi}_{ip}^{(q)}
		n
		\Big)
		\nsp\psp,
	\end{split}
\end{equation}
and ${\gamma}_p^{}[n]$ is 
additive white Gaussian noise. Here, $\alpha_{ip}^{(q)}$, $l_{ip}^{(q)}$ and ${\xi}_{ip}^{(q)}$ are, respectively, the unknown (complex) amplitude, the time delay and the (normalized) Doppler shift encountered by the $i$-th transmit waveform collected by the $p$-th receiver after reflection at the $q$-th target.
With respect to 
noise, the target echo signals are boosted by the processing gain resulting from the correlation.
The targets are detected, and TD and DS are estimated through range-Doppler correlation; each receiver contains ${M}$ correlation filters,  
each corresponding to a transmit sequence.
%
%
The received signal over ${\eta}{N}$ samples,  where ${\eta}\in\mathbb{N}\pspp$, 
is cross-correlated with the frequency-shifted ${\eta}$ replicas of transmit sequences 
(where the frequency shifts correspond to hypothesized values for the Doppler in the received signal) 
to produce the range-Doppler map for a coherent integration interval; over each such map, target detections are performed.

\section{SC Algorithm Using DCFT}\label{sec:algorithm}


When the received signal is composed of reflected echoes of various transmit signals, the targets resulting in weak reflections cannot be detected when the cross-correlation side lobes corresponding to stronger target reflections are larger  than the compression gain from autocorrelation of the weaker reflection. 
This scenario occurs often in practice when a target  has relatively low RCS or when the target is very far from the transmitter and receiver relative to some other target yet within the coverage area.
We propose a solution to address this problem in multi-transmitter DRS using ZC sequences. 


For an even-length, time-delayed ZC sequence $z_{u,N}^{}[n-l]\pspp$, ${N}\in 2\mathbb{N}\pspp$, it can be seen that when the chirp rate ${\beta}$ is matched to the seed of the sequence, 
i.e., 
${\beta}=u\pspp$,
the DCFT is reduced to the DFT 
and the DCFT results in a peak 
of magnitude $\sqrt{{N}}$ 
at
${k}=\modulo{ul}{{N}}$.  
The magnitude of DCFT under the unmatched case is bounded by a value, which 
will be gracefully larger than 
$1\pspp$, and this shall be proven by observing the relation between ZC sequences' PACFs and PCCFs and their DCFTs and proceeding 
similarly to \cite{kang2011generalized}.
%
%
By utilizing this property, 
a new algorithm to detect multiple targets based on successive cancellations in the DCFT domain is proposed.

\begin{algorithm}[h]
	\caption{{Multi-Target Detection via Successive Cancellation in the DCFT Domain - SC\hyp DCFT.}}
	\label{alg:sic}
	\textbf{Input}: Received signal at the $p$-th receiver 
	$v_{p}^{}[n]\pspp$.
	\\
	\textbf{Initialization}: Initialize the residual signal, $r_p^{}[n] = v_{p}^{}[n]\pspp$.
	\begin{algorithmic}[1]
		\Repeat
		
		\For{$i=1,\,2,\,\ldots,\,M$}\vspace{-0.25em}
		\State 
		\hspace{-0.65em}\begin{tabular}{L{20em}}
			Compute the range-Doppler map 
			through ambiguity function between $r_p^{}[n]$ and $s_i^{}[n]\pspp$.
		\end{tabular}
		\vspace{0.05em}
		\For{each detection $\kappa$ in the range-Doppler map, } \vspace{-0.8em}
		\State Estimate the delay $l_{ip}^{(\kappa)}$ and Doppler shift $\xi_{ip}^{(\kappa)}$. \vspace{-0.35em}
		\State $g_p^{}[k]=\operatorname{DCFT}_{\beta=u_i^{}}^{}\!\Big\{ r_p^{}[n] \exp\nsp\Big(\nsp-\imag 2\pi {\xi}_{ip}^{(\kappa)}
		\Big)\Big\}$. \vspace{-0.35em}
		\State $g_p^{}\big[\bigmodulo{u\psp l_{ip}^{(\kappa)}}{N}\big]\gets 0\pspp$. \vspace{0.5em}
		\State $r_p^{}[n]\gets\operatorname{IDCFT}_{\beta=u_i^{}}^{}\{ g_p^{}[k]\} \times \exp\nsp\Big(\imag 2\pi {\xi}_{ip}^{(\kappa)}
		\Big)$. \vspace{-0.6em}
		\EndFor
		\EndFor
		\Until{no further targets are detected.}
		%
	\end{algorithmic}
	\textbf{Output}: Detected targets and their TD and DS measurements.
\end{algorithm}
A reflection of a ZC sequence present in the composite received signal will get transformed into an impulse in the DCFT domain when $\beta$ is matched to the seed of the sequence. We propose nulling such impulses to remove the interference; once the inverse transform is applied, the corresponding ZC sequence would have been removed in the time domain. 
After such a cancellation, any other weaker reflection present in the residue might be visible, as removing the stronger ZC sequence also removes the corresponding side lobes in the  subsequent range-Doppler maps. 
This renders the detection of targets which result in weak reflections possible. By successively performing such cancellations in the DCFT domain, more targets which were originally undetectable can now be detected. 
This process can be repeated until no more detectable target reflections are 
in the residue.
	%

However, in the presence of Doppler, ideal behaviour is gradually lost, and a ZC sequence will not get transformed into an impulse by the DCFT but rather into a Dirichlet (periodic) sinc (see \cite[eq. (17)]{hua2014analysis}). Also, the increase in Doppler shift results in a rapid decrease of the peak. 
%
%
This presents a challenge, 
and 
the SC has to be performed after appropriate handling of the Doppler.
%
%
%
Since the Doppler of the detected target can be estimated from the range-Doppler map, the entire signal can be de-rotated in order to minimize the Doppler present in the strong reflection from a detected target. 
Under the DCFT, such a strong replica of the ZC sequence with a low residual Doppler would  become an impulse and can be filtered. After performing this step, the residues can be rotated back to remove the Doppler, which was injected into the weaker reflections when de-rotating the strong reflections. By proceeding this way, the targets resulting in weaker echoes can be detected through SC.
%
%
%
%
\commreplace{\\{}\indent$\!\!\!$}{}
The proposed scheme  is summarized in Algorithm \ref{alg:sic}.
\begin{figure}[ht]
	\vskip 0.05in
	\begin{center}
		\centerline{\includegraphics[width=0.97\columnwidth]{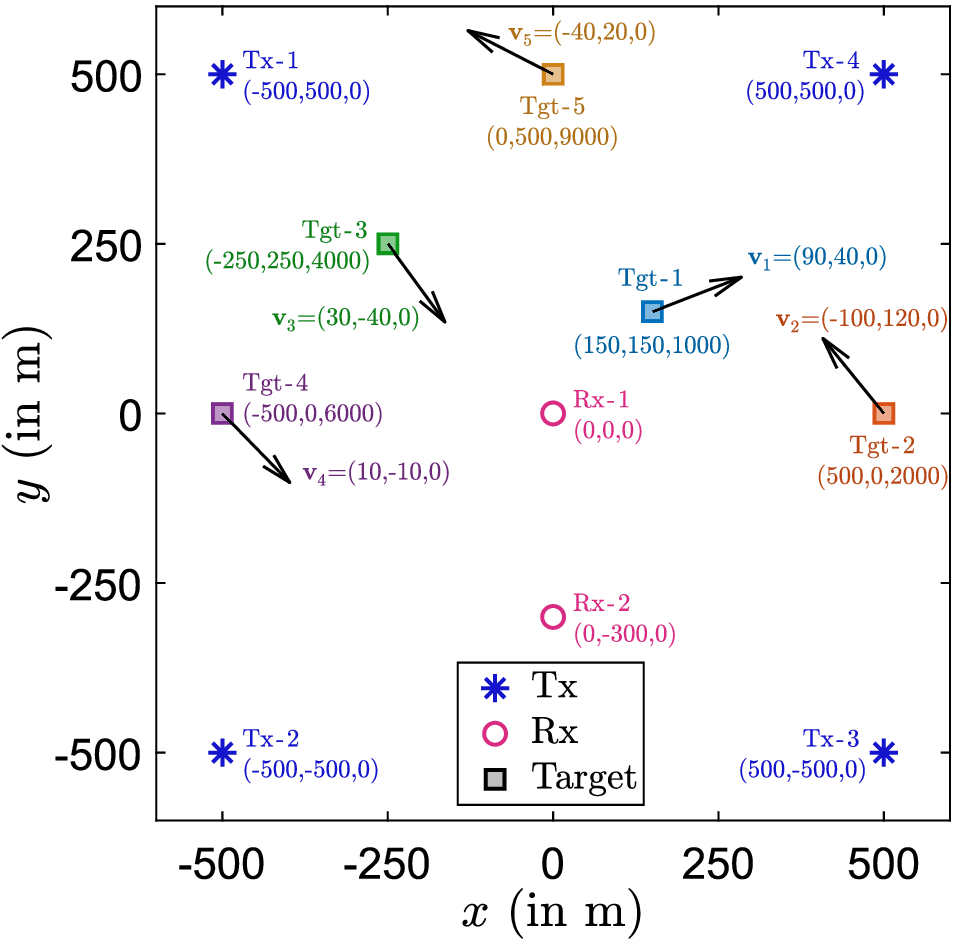}}
		\caption{Top view of the system geometry of the simulated DRS. }
		\label{fig:tx_rx_tgt_loc}
	\end{center}
	\vskip -0.05in
\end{figure}
\begin{figure*}[h!]
	\begin{subfigure}{0.49\linewidth}
		\centering
		\includegraphics[width=0.95\linewidth]{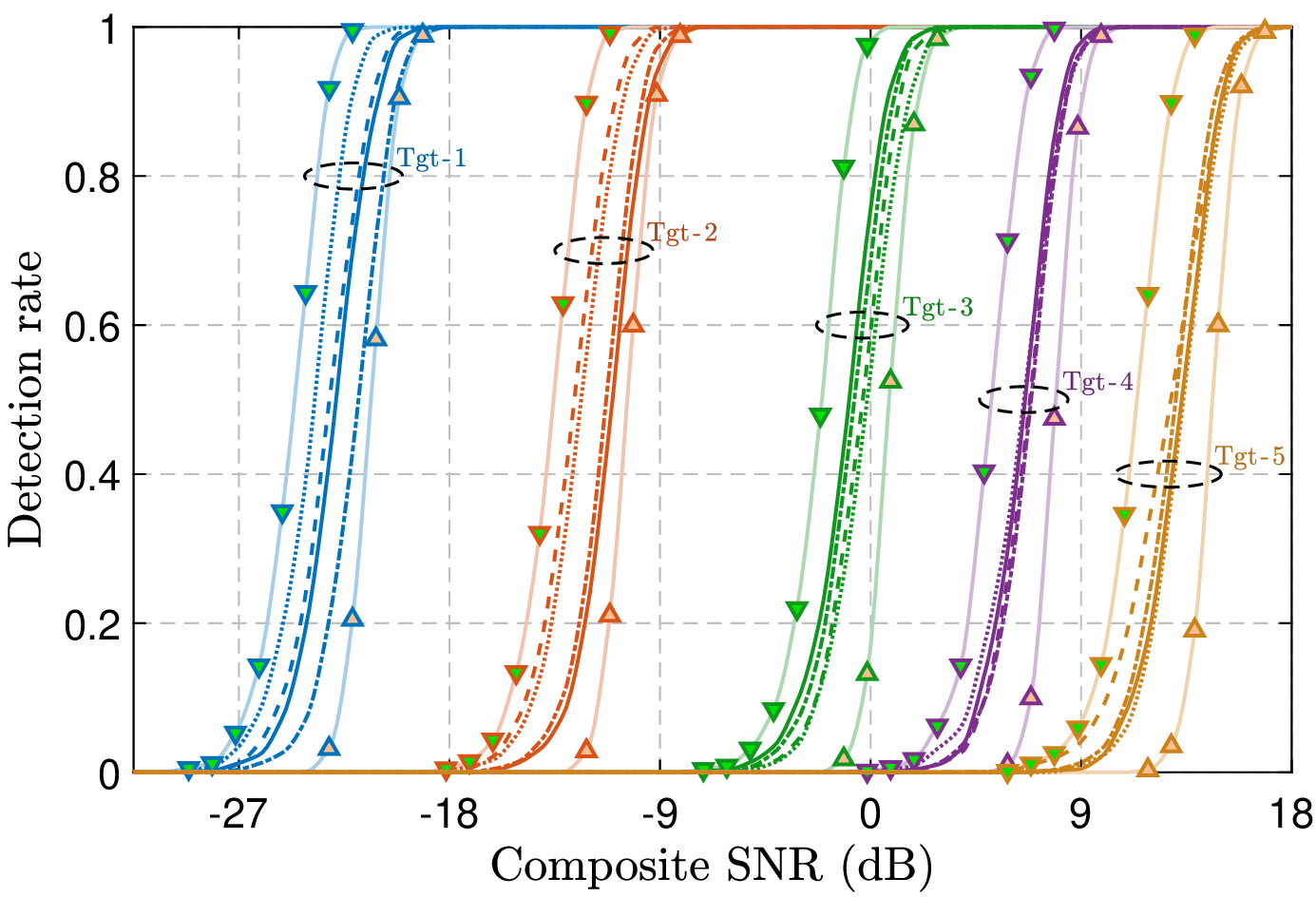}\vspace{-0.5ex}
		\caption{first receiver - SC\hyp DCFT.}
		\label{fig:case_1}
	\end{subfigure}
	\hspace{0.02\linewidth}
	\begin{subfigure}{0.49\linewidth}
		\centering
		\includegraphics[width=0.95\linewidth]{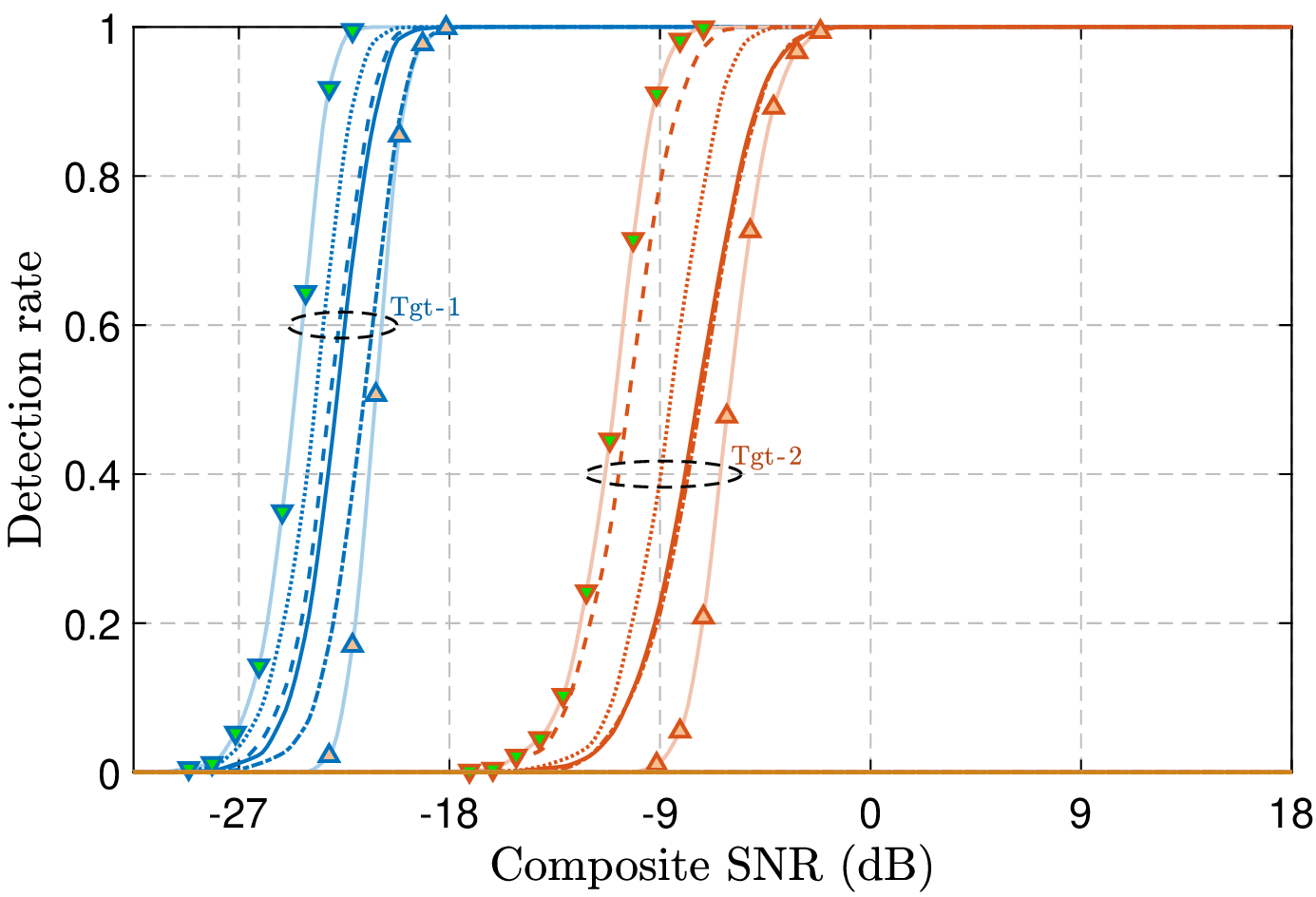}\vspace{-0.5ex}
		\caption{first receiver - without SC.}
		\label{fig:case_1_no_SIC}
	\end{subfigure}
	\vskip 2.2ex
	\begin{subfigure}{0.49\linewidth}
		\centering
		\includegraphics[width=0.95\linewidth]{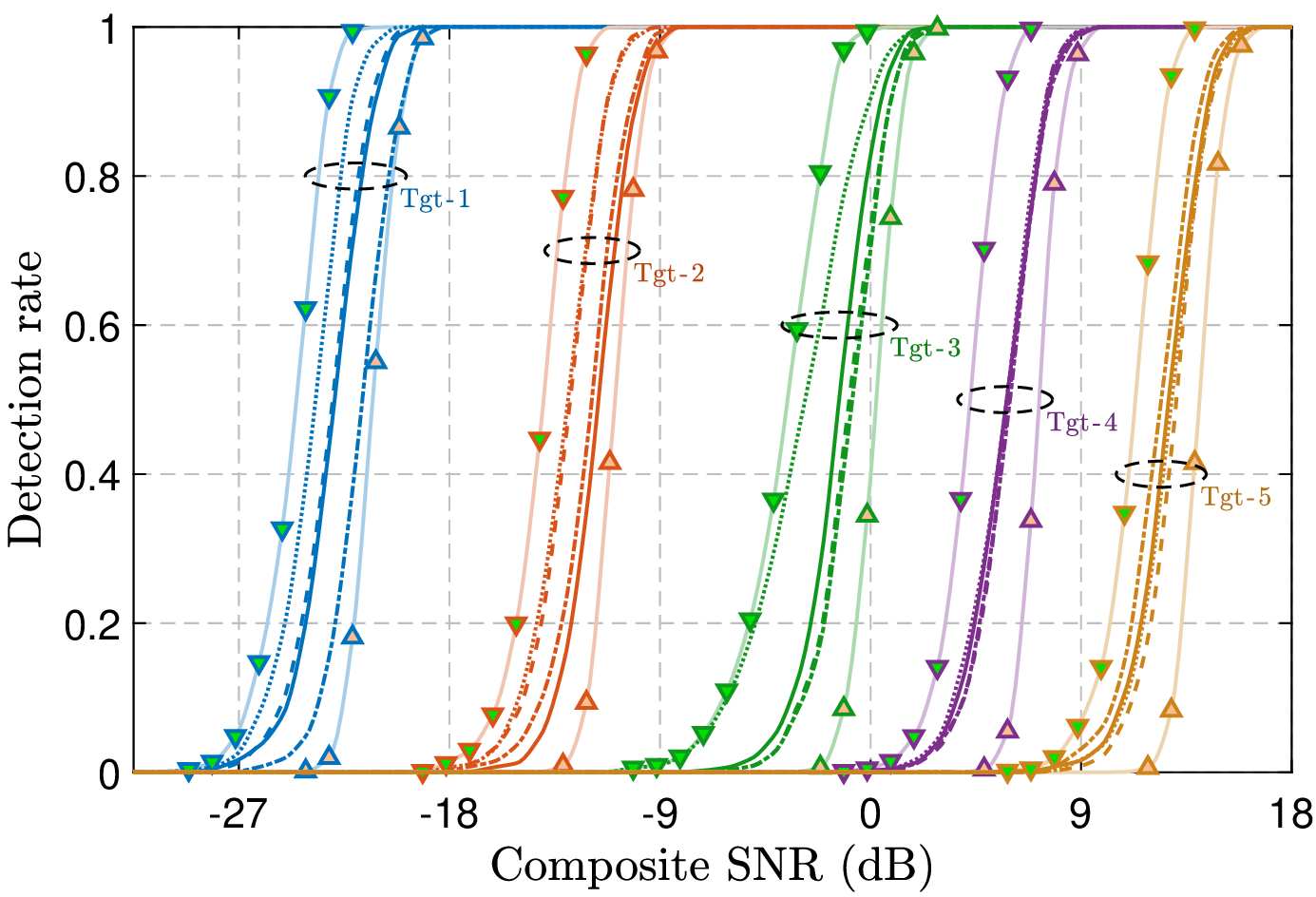}\vspace{-0.5ex}
		\caption{second receiver - SC\hyp DCFT.\vspace{1ex}}
		\label{fig:case_2}
	\end{subfigure}
	\hspace{0.02\linewidth}
	\begin{subfigure}{0.49\linewidth}
		\centering
		\includegraphics[width=0.95\linewidth]{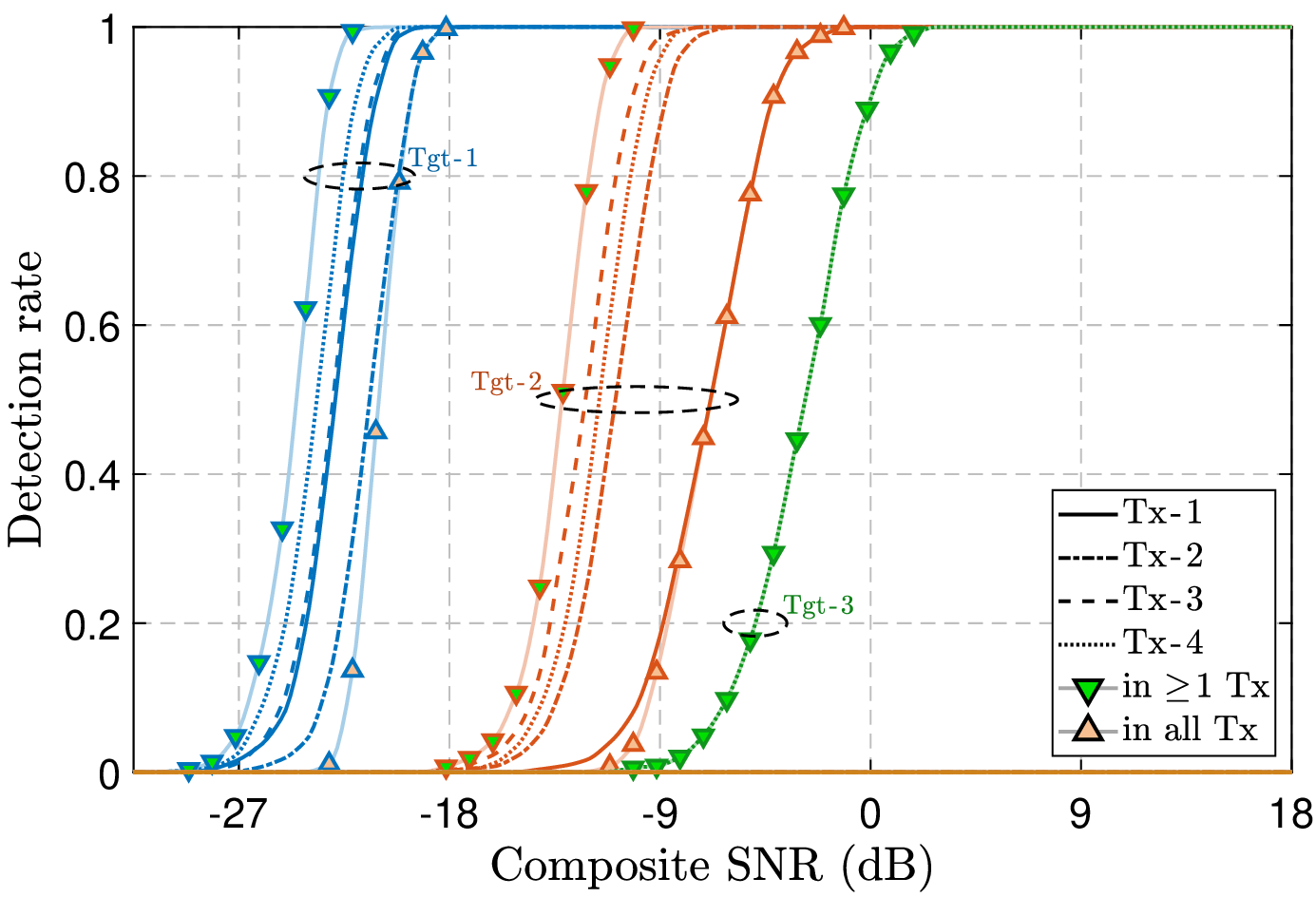}\vspace{-0.5ex}
		\caption{second receiver - without SC.\vspace{1ex}}
		\label{fig:case_2_no_SIC}
	\end{subfigure}
	\vskip 0.2ex
	\caption{Detection rates of the targets for Case 1.}
	\label{fig:case_1_and_2}
\end{figure*}

\section{Simulation Results}\label{sec:simulation}

Computer simulations are performed to validate the proposed solution. We consider a VHF DRS operating on a $20\psp $MHz band centered at $500\psp $MHz. The system employs several CW transmitters which transmit ZC sequences of length ${N}=2048\pspp$, resulting in a pulse duration of $102.4\psp\mu$s; coherent integration of received waveform is performed over ${\eta}=8$ pulses, resulting in a coherent integration time of $819.2\psp\mu$s.
%
We mostly consider the system configuration presented in 
Figure \ref{fig:tx_rx_tgt_loc}, where positions are marked in metres and velocities are marked in m/s. 
There are four transmitters (Tx\hyp1\psp:\psp Tx\hyp4) and two receivers (Rx\hyp1 and Rx\hyp2), which are ground-based and stationary. All the transmitters radiate 
with the same power of $500\psp$W. 
There are five air-borne targets (Tgt\hyp1\psp:\psp Tgt\hyp5) that are moving with constant velocities (labelled as $\mbf{v}_1^{}$\pspp:\pspp $\mbf{v}_5^{}$). 
All the targets in the figure have unit RCS. The delays, Doppler shifts and attenuations of the received waveforms in \eqref{eq:rx} are accordingly determined from this system configuration.
The performance is studied under four different cases.
\begin{enumerate}[label=$\bullet$]
	
	\item \textbf{Case 1:} All four transmitters 
	are radiating, and all five targets are in the surveillance area. 
	
	\item \textbf{Case 2:} All  the targets are in the surveillance area, 
	but only two transmitters (Tx\hyp1 and  Tx\hyp3) radiate.
	
	\item \textbf{Case 3:} All the transmitters 
	are radiating, 
	but only two targets (Tgt\hyp2 and Tgt\hyp4) are present in the area.
	
	\item \textbf{Case 4:} All the transmitters 
	are radiating, 
	and two targets (Tgt\hyp3 and Tgt\hyp6) are present. Tgt\hyp6 (not marked in the figure, located at $(-500,0,3975)$, $\mbf{v}_6^{}=\mbf{v}_4^{}$) has an altitude very close to that of Tgt\hyp3 and has a low RCS of \pspp $\pspp0.025\psp\text{m}^2_{}$,  16dB \pspp below \pspp that \pspp of \pspp Tgt\hyp3, \pspp
	and \pspp results \pspp in colliding peaks in two of the four matched filters in Rx\hyp1. 
\end{enumerate}
\begin{figure*}[h!]
	\begin{subfigure}{0.49\linewidth}
		\centering
		\includegraphics[width=0.95\linewidth]{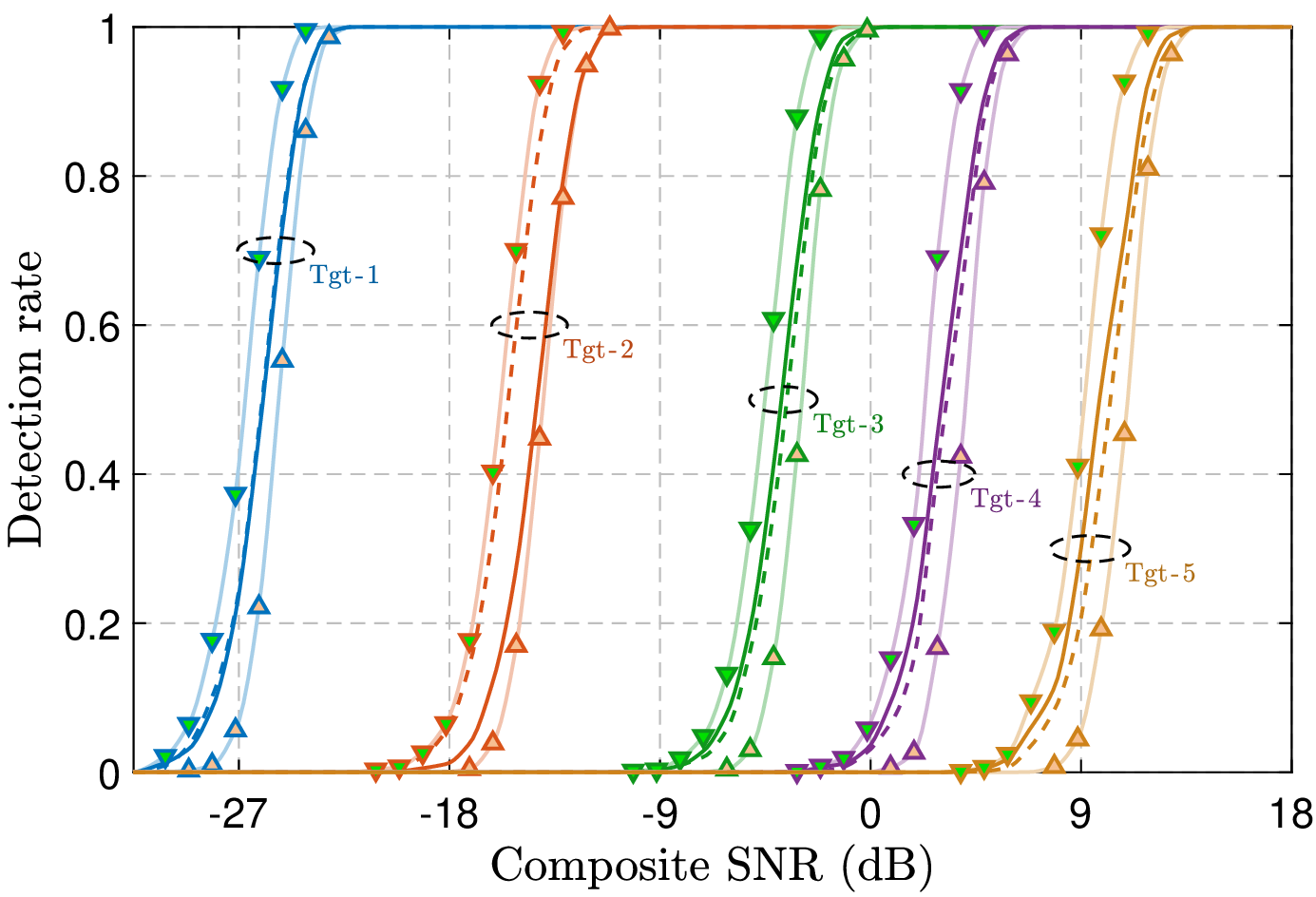}\vspace{-0.5ex}
		\caption{}
		\label{fig:lesser_tx}
	\end{subfigure}
	\hspace{0.02\linewidth}
	\begin{subfigure}{0.49\linewidth}
		\centering
		\includegraphics[width=0.95\linewidth]{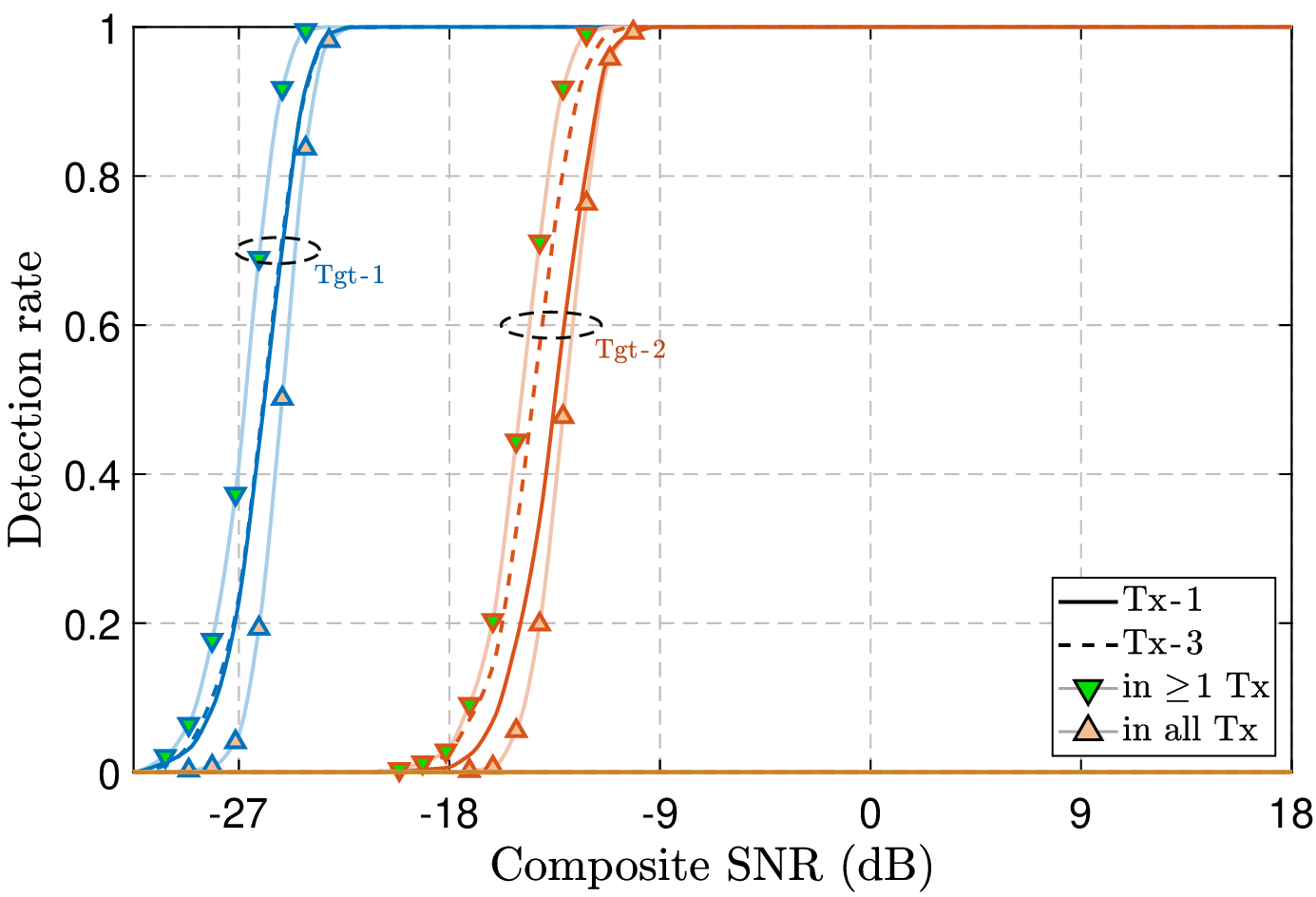}\vspace{-0.5ex}
		\caption{}
		\label{fig:lesser_tx_no_SIC}
	\end{subfigure}
	\vskip 0.2ex
	\caption{Detection rates of the targets in Rx\hyp1 for Case 2, (a) SC\hyp DCFT (b) without SC.}
	\label{fig:lesser_tx_all}
\end{figure*}
\begin{figure*}[h!]
	\begin{subfigure}{0.49\linewidth}
		\centering
		\includegraphics[width=0.95\linewidth]{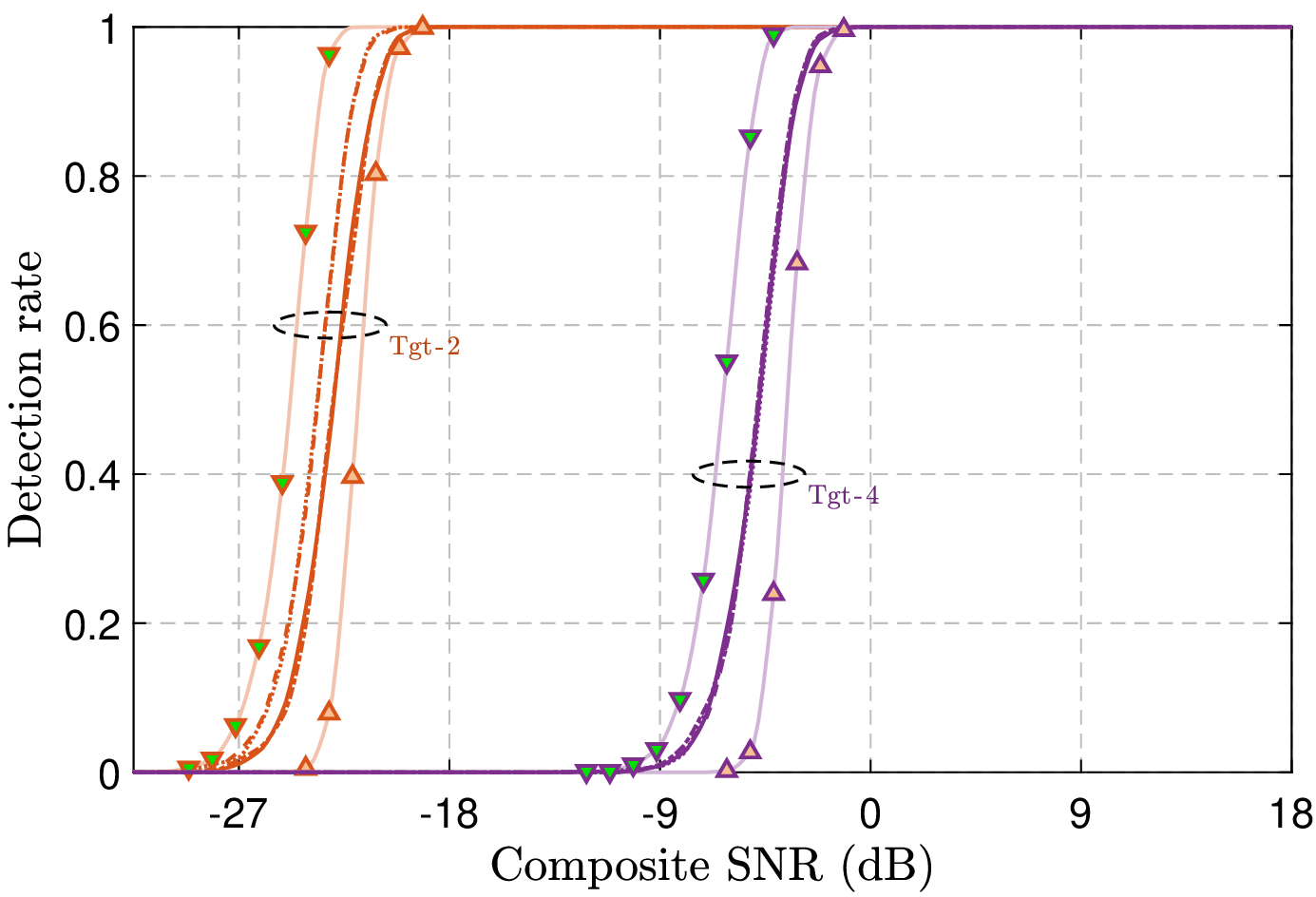}\vspace{-0.5ex}
		\caption{}
		\label{fig:lesser_tgt}
	\end{subfigure}
	\hspace{0.02\linewidth}
	\begin{subfigure}{0.49\linewidth}
		\centering
		\includegraphics[width=0.95\linewidth]{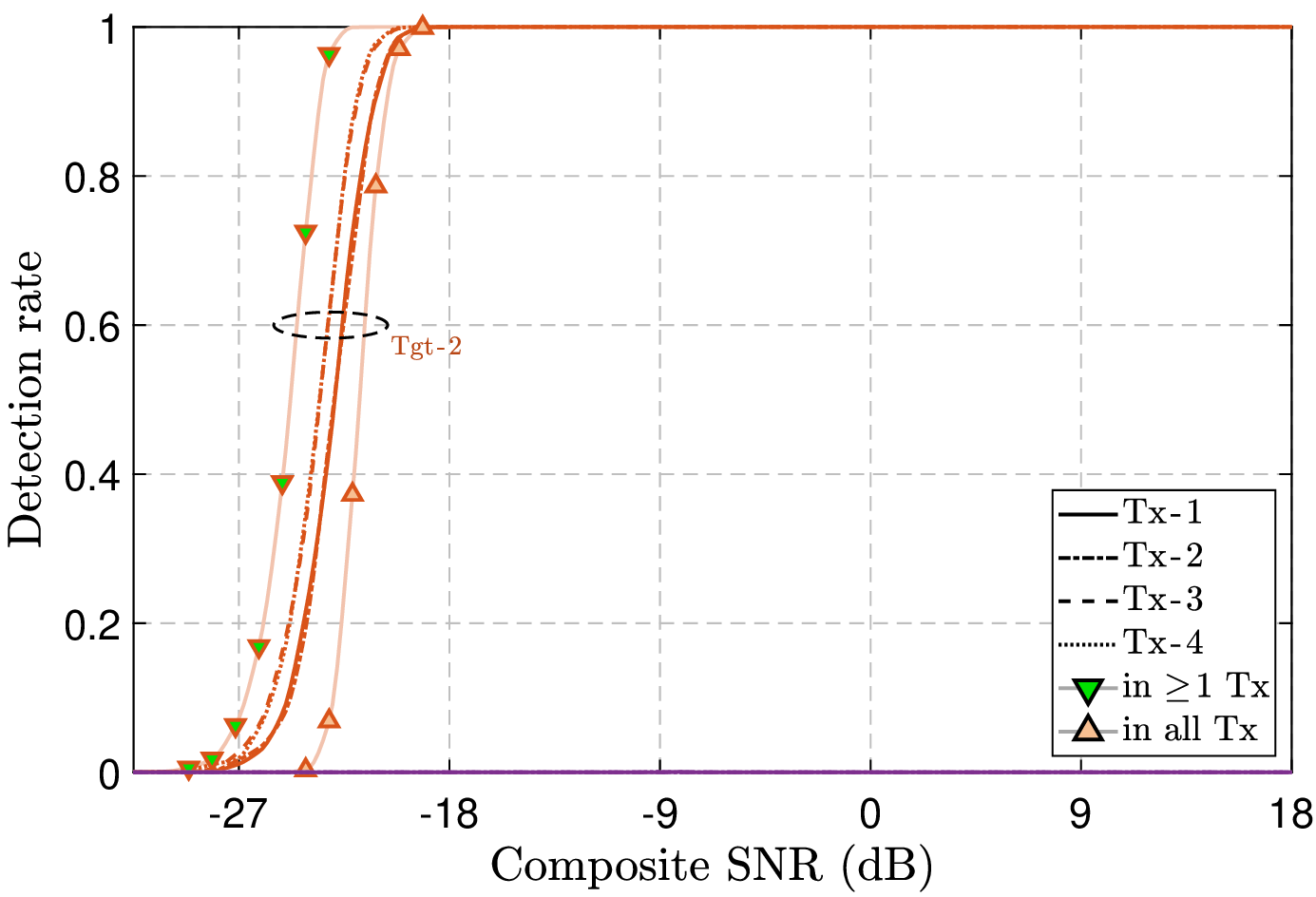}\vspace{-0.5ex}
		\caption{}
		\label{fig:lesser_tgt_no_SIC}
	\end{subfigure}
	\vskip 0.2ex
	\caption{Detection rates of the targets in Rx\hyp1 for Case 3, (a) SC\hyp DCFT (b) without SC.}
	\label{fig:lesser_tgt_all}
\end{figure*}

\begin{figure*}[h!]
	\begin{subfigure}{0.49\linewidth}
		\centering
		\includegraphics[width=0.95\linewidth]{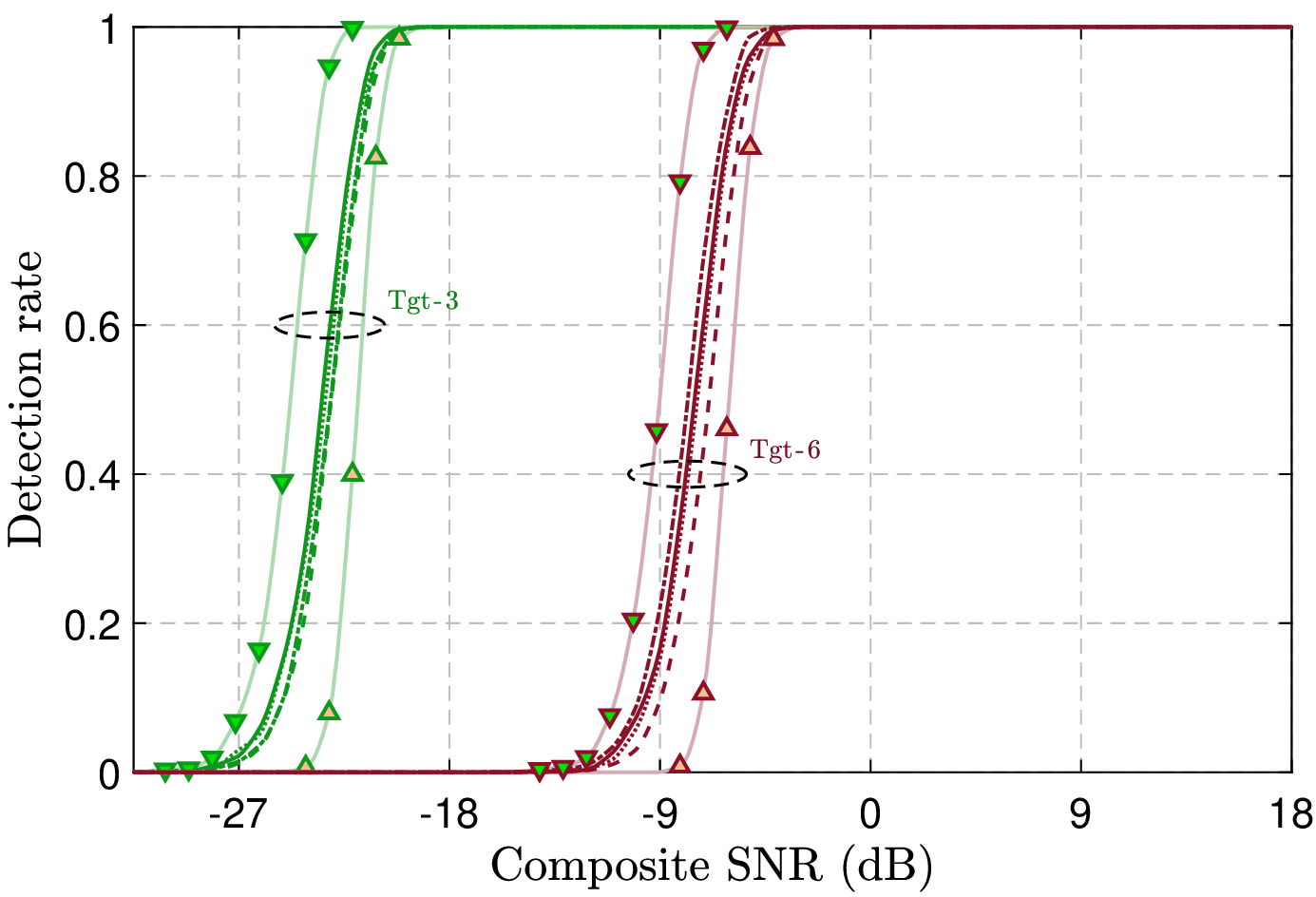}\vspace{-0.5ex}
		\caption{}
		\label{fig:adjacent_peaks}
	\end{subfigure}
	\hspace{0.02\linewidth}
	\begin{subfigure}{0.49\linewidth}
		\centering
		\includegraphics[width=0.95\linewidth]{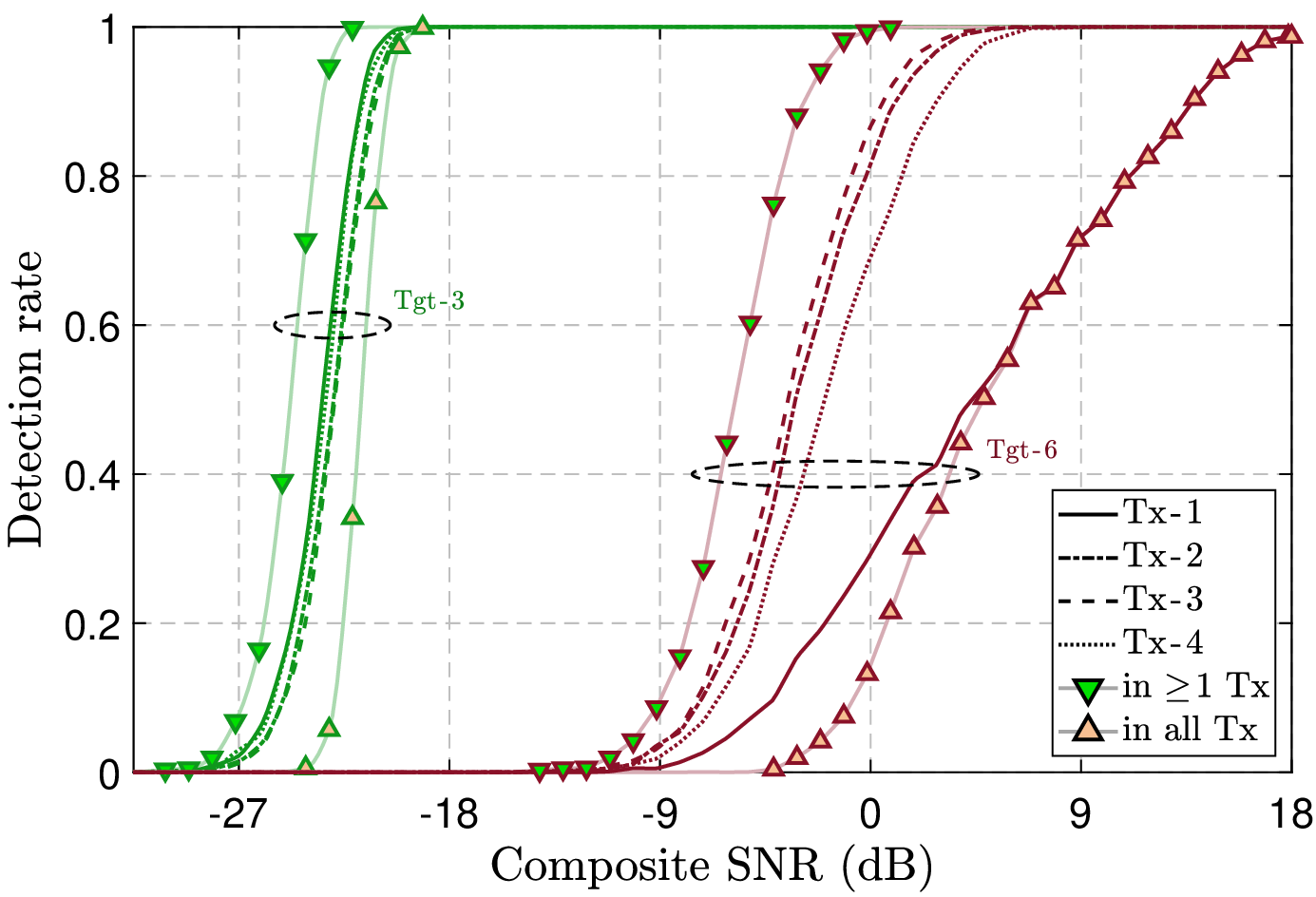}\vspace{-0.5ex}
		\caption{}
		\label{fig:adjacent_peaks_no_SIC}
	\end{subfigure}
	\vskip 0.2ex
	\caption{{Detection rates of the targets in Rx\hyp1 for Case 4, (a) SC\hyp DCFT (b) without SC}}
	\label{fig:adjacent_peaks_all}
\end{figure*}

We study the detection rate (i.e., the fraction of times a target gets correctly)  as a function of composite Signal-to-Noise-Ratio (SNR) at the receivers, which is the ratio of powers of composite received signal, ${v}_{p}^{(0)}[n]\pspp$, and the noise, ${\gamma}_p^{}[n]\pspp$, in \eqref{eq:rx} and consider $2000$ Monte-Carlo trials. The detection rates of each target in each of the transmit waveforms are individually shown, and the rates corresponding to detection in at least one waveform and all the waveforms are also studied.

Results corresponding to Case 1 are presented in Figure  \ref{fig:case_1_and_2} and Table \ref{tab:SC_time}. 
Figures  \ref{fig:case_1} and \ref{fig:case_1_no_SIC} show the detection rates of the proposed approach (SC\hyp DCFT in Algorithm \ref{alg:sic}) and those of raw detection in Rx\hyp1. The corresponding results for Rx\hyp2 are provided in Figures \ref{fig:case_2} and \ref{fig:case_2_no_SIC}. From the plots, it is evident that in both the receivers, the raw detection can detect only two low-flying targets (Tgt\hyp1 and Tgt\hyp6) in all the transmit waveforms; additionally, in Rx\hyp2,  Tgt\hyp3 at the altitude of $4\psp$km gets detected in the waveform of Tx\hyp4. Meanwhile, the proposed scheme results in the \textit{detection of all the targets in all the transmit waveforms}. In addition, the detection rates for Tgt\hyp2 are significantly better than the raw detection; SC\hyp DCFT detects Tgt\hyp2 in all the transmit waveforms 
at the composite SNR of $-7\psp$dB, which is not the case with raw detection. 
The raw detection method could not detect since its performance is heavily limited by interference due to reflections from Tgt\hyp1 and Tgt\hyp2. The performance of  the proposed scheme is mainly \textit{limited by noise}, and 
the detections of Tgt\hyp3\psp:\psp Tgt\hyp5 occur at high composite SNRs just because of the weak contributions of corresponding reflections to the composite signal  ${v}_{p}^{(0)}[n]\pspp$. 

In Table \ref{tab:SC_time}, the detection performance of our scheme is compared with that of the conventional successive cancellation in the time domain, where  the time domain replicas of the detected waveforms are reconstructed from estimated amplitude, delay and Doppler estimates from the range-Doppler map (as in \eqref{eq:rx}), and subtracted from the received signal to further detect other targets from the residue. We study the detection rates of the highest target, Tgt\hyp5, in the waveform of TX\hyp3 in Rx\hyp1. Though SC in the time domain, unlike the raw detection, can detect Tgt\hyp5, the proposed scheme outperforms it with better detection rates.
These results show the efficacy of the proposed method in handling multiple targets by properly handling interferences from strong reflections, limiting the performance only to the noise level.


\begin{table}[h!] 
	\caption{Detection rates of Target-5 in Case 1 with respect to the third transmitter in the first receiver }
	\label{tab:SC_time}
	\setlength{\tabcolsep}{3.4pt}
	\begin{center}
		\begin{tabular}{?c?c|c|c|c|c|c?} 
			
			\clineB{1-7}{2}
			%
			\begin{tabular}{C{12ex}}
				\vspace{-4.75ex}	 \textbf{Composite} \\[-2.5ex] \textbf{SNR (dB)}
			\end{tabular} &
			$9$ & $10$ & $11$ & $12$ & $13$ & $14$ \\[1ex] 
			\clineB{1-7}{2}
			\specialrule{1pt}{2pt}{0pt}	
			
			\begin{tabular}{C{12ex}}
				\vspace{-4.75ex} \textbf{Proposed} \\[-2.75ex] \textbf{{SC\hyp DCFT}}
			\end{tabular} &
			$0.0370$ & $0.0835$ & $0.1745$ & $0.3305$ & $0.5030$ & $0.7050$ \\
			\clineB{1-7}{1}
			
			\begin{tabular}{C{12ex}}
				\vspace{-4.75ex} \textbf{SC in} \\[-2.75ex] $\!\!\nsp$\textbf{Time Domain}$\!\!\!$
			\end{tabular} &
			$0.0040$ & $0.0195$ & $0.0650$ & $0.1810$ & $0.3995$ & $0.6680$ \\ 
			\specialrule{1pt}{0pt}{0pt}
			
		\end{tabular}
	\end{center}
\end{table}

In Figure  \ref{fig:lesser_tx_all}, the detection performance in Rx\hyp1 is presented for the case with a lesser number of transmitters than before (Case 2), and in Figure  \ref{fig:lesser_tgt_all}, it is plotted for a scenario with a lesser number of transmitters (Case 3). Both cases offer lesser overall interference than before. Despite that, raw detection could not detect any targets which were undetected in the previous study. The proposed approach is powerful and detects all the targets in all the waveforms, and it also offers performance improvement in these cases where the interference is less,  which is evident by comparing  Figures  \ref{fig:lesser_tx} and \ref{fig:lesser_tgt} with Figure  \ref{fig:case_1}. For the challenging Case 4, the results are plotted in Figure  \ref{fig:adjacent_peaks_all}.  It can be observed again that the proposed method offers better performance than the raw detector, particularly for the detection of Tgt\hyp6.
\begin{figure}[ht]
	\begin{center}
		\centerline{\includegraphics[width=0.95\columnwidth]{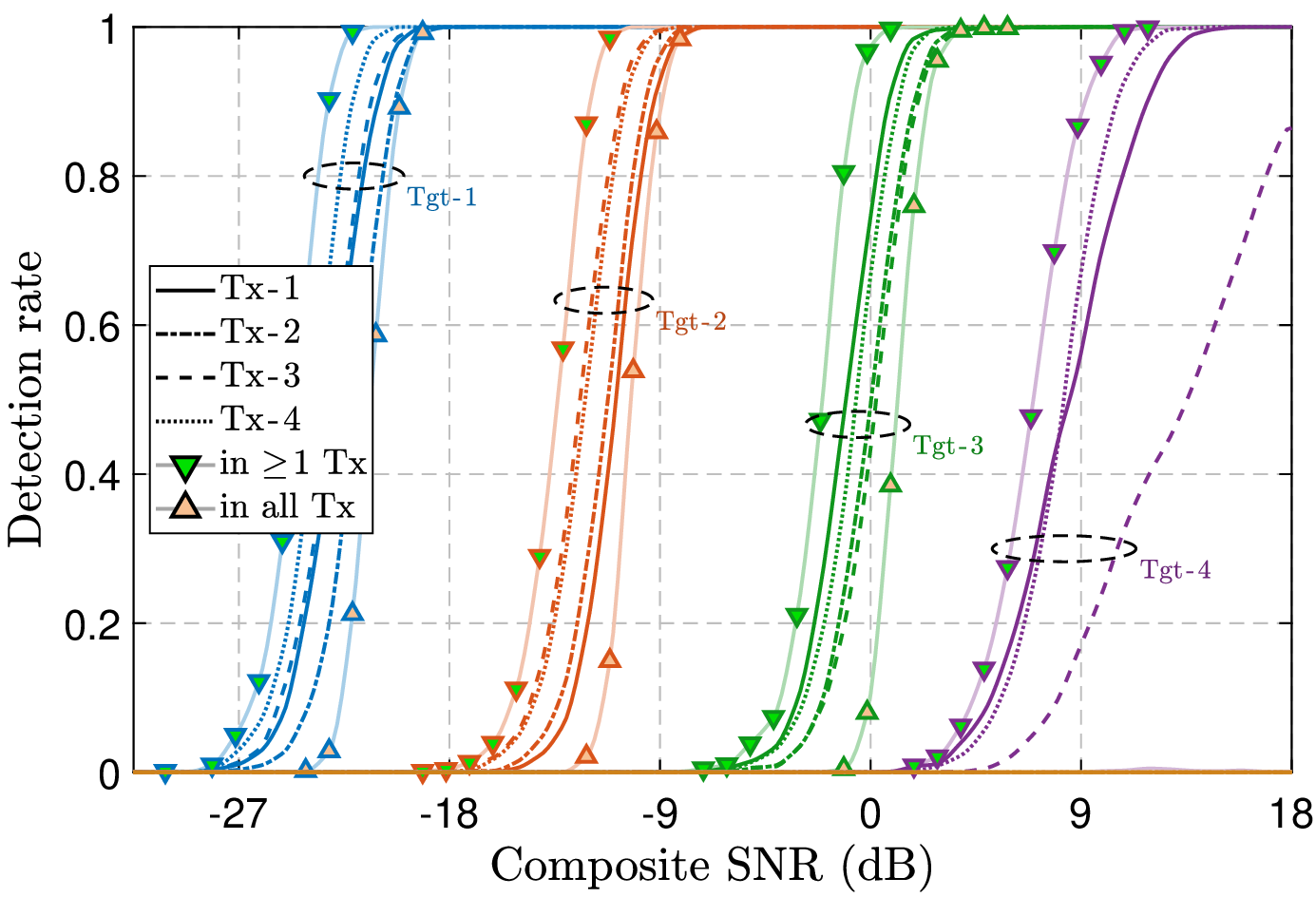}}
		\caption{Detection rates of the targets under band limited waveforms in Rx\hyp1 for Case 1 .}
		\label{fig:case_1_band_limit}
	\end{center}
\end{figure}
Finally, in Figure  \ref{fig:case_1_band_limit}, we study the performance of the proposed method for the case when waveforms are band-limited to $90\%$ of the available band. 
In this scenario, the transmit signals will not be a perfect ZC sequence; hence, performance deterioration is expected. 
However, the performance is still good, detecting up to four targets. Future work shall focus on improving the performance further in such practical settings.


\section{Conclusions}\label{sec:conclusion}
In this work, we have described a novel  
multi-target detection scheme based on successive cancellations 
for 
continuous wave, distributed radar systems 
functioning with several transmitters that 
radiate  Zadoff-Chu  sequences in the same band. 
%
%
%
%
In the receivers, successive cancellations are performed using the Discrete Chirp-Fourier Transform after carefully compensating for the Doppler shifts. 
This minimizes the interference 
to other weak target reflections in the received signal, 
which get detected subsequently. 
%
%
Through extensive numerical simulations, 
the effectiveness of the proposed method in detecting multiple targets is demonstrated in a wide range of conditions.

\section*{Acknowledgment}
\addcontentsline{toc}{section}{Acknowledgment}

The authors would like to sincerely 
thank Mr. Krishna Madan Yelamarty, PhD scholar in the TelWiSe Group at 
IIT Madras, for the 
many useful discussions 
on this work.
%
%

{
	\bibliographystyle{IEEEtran}
	\bibliography{ref_radar_detec}
}

\end{document}